\documentclass{ws-procs10x7}

\begin{document}

\title{Searches for new Physics at HERA }

\author{A. Sch\"oning}

\address{(on behalf of the H1 and ZEUS collaborations) \\
Institute of Particle Physics, ETH Z\"urich, ETH H\"onggerberg, CH-8093 Z\"urich\\E-mail: schoning@mail.desy.de}

\twocolumn[ \maketitle \abstract{Results of a general search for
    new phenomena at high transverse momentum and of a dedicated search for events with isolated leptons and missing 
transverse momentum are reported. 
These searches were performed on a data sample collected at HERA in $e^\pm p$ 
collisions with the H1 detector in the 
period 1994--2004. The results obtained in the isolated lepton search are 
compared with those obtained by the ZEUS collaboration in the period 1994--2000
 and limits on the anomalous FCNC production of single top events are presented.}
]

\section{Introduction}
The high-energy frontiers of accelerators serve as ideal hunting grounds
to look for new phenomena and physics beyond the Standard Model (SM). 
A general search for new phenomena at high transverse momentum was recently
performed in a model independent framework  
by the H1 Collaboration~\cite{ref:generic1}. 
The analysis used $e^\pm p$ data  taken at center of mass energies
$s^{{1}/{2}}$= 300, 320~GeV in the period 1994-2000 corresponding to an integrated luminosity of ${\cal{L}}=118$~pb$^{-1}$
and was repeated with recent data~\cite{ref:hera2search} taken after the HERA luminosity upgrade 
in 2003/2004 
corresponding to ${\cal{L}}=45$~pb$^{-1}$ ($s^{1/2}$= 320~GeV). 
Results are presented in section~\ref{sec:generic}.

The excess of events with isolated electrons or muons and missing 
transverse momentum ($p_T$) observed in the period 1994-2000 by the H1 collaboration \cite{ref:h1isol,ref:h1tau}
has motivated a repeat of the same analysis on the newest data
 taken in 2003/2004.
In section~\ref{sec:isol}
the event yields are compared 
with results obtained by the ZEUS collaboration~\cite{ref:zeusisol,ref:zeustau} performed on data taken from 1994-2000 (${\cal{L}}=130$~pb$^{-1}$) 
where a slight excess of tau 
events compared to the SM expectation was seen. 

In section~\ref{sec:top} searches for events from anomalously produced top quarks
are discussed, which in case of the semileptonic decay contribute to the
isolated lepton with missing $p_T$ signature and might explain the observed
excesses. Limits on the involved 
anomalous couplings are presented and compared with similar limits obtained
at other colliders.

\section{General search} \label{sec:generic}
\begin{figure}[thb]
\setlength{\unitlength}{1cm}
\begin{picture}(10,10.5)
     \put(-0.1,10.5){\epsfig{file=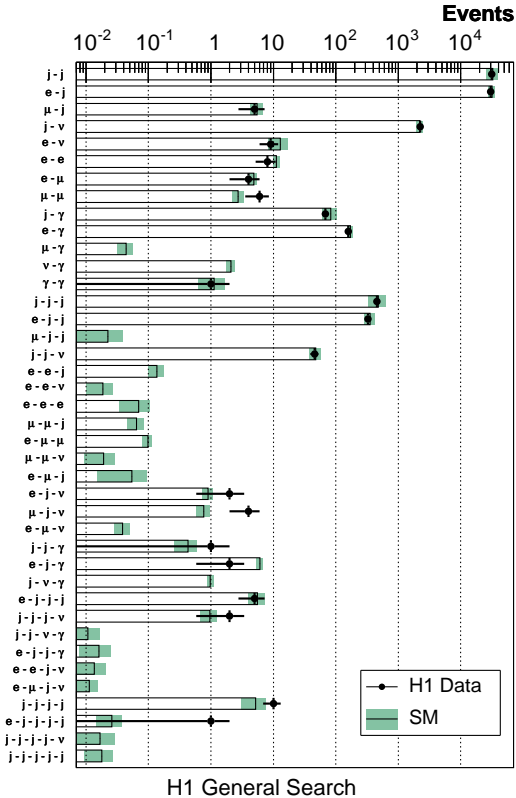,angle=-90,width=70mm}}
\end{picture}
\caption{HERA~I event yields in the general search for new phenomena at 
high $p_T$. Data (points) are compared to the SM expectation
(histogram bars). The uncertainty of the SM prediction is indicated by
the shaded band.}
\label{fig:generic_class1}
\end{figure}

The H1 collaboration has performed
a general search for new phenomena 
by looking for deviation from the SM prediction at high transverse
momentum~\cite{ref:generic1}. 
For the first time all event topologies involving objects like
electrons ($e$), photons ($\gamma$), muons ($\mu$), neutrinos ($\nu=$~missing particles) and jets ($j$) are 
investigated in a single analysis.
Event classes are defined consisting of at least two clearly
identified and isolated objects $i$ with a minimum transverse momentum of 
$p_T^i>20$~GeV.
Events were found in 22 classes. The event yields
span several orders of magnitude, see fig~\ref{fig:generic_class1}. 
Overall good agreement
with the SM prediction was found.

In order to enhance sensitivity to new physics processes differential
 distributions as function of the total invariant mass of all objects
 $M_{\rm All}$ and of the summed transverse momentum $\sum_i{p_T^i}$ 
 were investigated.
Possible deviations (excess or deficit) were searched for with a new
 algorithm and statistically
 quantified.
After studying all event classes the most significant deviation was 
found in the $\mu j \nu$ channel, where the main SM background comes
 from single W-production.
The probability of finding in the H1 data an excess in any of the event classes more 
significant than the observed excess in the  $\mu j \nu$ class, was
 found to be 3\% for the $\sum_i{p_T^i}$ distribution and 28\% in the
 $M_{\rm All}$ distribution.

The same analysis was repeated with the recent data taken by the H1
experiment in 2003/2004 based on an integrated luminosity of
45~pb$^{-1}$.
Again good overall agreement between data and the SM expectation was
found.
In this dataset
the most significant deviation was found in the $e
j \nu$ channel, to which also single W-production mainly contributes
in the SM.
The excesses observed in the $e j \nu$ and $\mu j \nu$ classes are
further detailed in the next section.

\section{Isolated leptons with missing transverse momentum} \label{sec:isol}

\begin{table*}
\caption{HERA I event yields in the search for isolated leptons with missing 
transverse momentum. The numbers are given for the electron, muon and tau channel for different
cuts $p_T^X$.}
\label{tab:iso_hera1}
\begin{tabular}{|c|c|c|c|}     
\hline
{HERA~I 1994-2000}   & \multicolumn{3}{c|}{observed/expected}
\\ \hline \hline
{{\bf H1} ${\cal L}(e^\pm p)=118$~pb$^{-1}$} & Electron & Muon &  Tau \\ \hline
 Full sample 
 & { 11 / 11.5 \small $\pm $1.5  } 
 & { { 8}  / 2.94 \small $\pm$ 0.51 }  
 & { 5  / 5.81 \small $\pm$ 1.36 }  
 \\
 $P_{T}^{X}>25$~GeV 
 & { {\bf 5} / 1.76 \small $\pm$ 0.29 } 
 & { {\bf 6} / 1.68 \small $\pm$ 0.30  } 
 & { 0 / 0.53 \small $\pm$ 0.10  } 
 \\ 
 $P_{T}^{X}>40$~GeV 
 & { { 3} / 0.66 \small $\pm$ 0.13 } 
 & { { 3} / 0.64 \small $\pm$ 0.14 } 
 & { 0 / 0.22 \small $\pm$ 0.05 } 
 \\ 
\hline
{{\bf ZEUS} ${\cal L}(e^\pm p)=130$~pb$^{-1}$} & Electron & Muon &  Tau \\ \hline
 Full sample 
 & { 24 / 20.6 \small $\pm$3.2  } 
 & { 12  / 11.9 \small $\pm$0.6 }  
 & { { 3}  / 0.4 \small $\pm$0.12  } 
  \\ 
 $P_{T}^{X}>25$~GeV 
 & { 2 / 2.9 \small $\pm$0.46 } 
 & { 5 / 2.75 \small $\pm$0.21  } 
 & { {\bf 2} / 0.2 \small $\pm$0.05  } 
\\ 
 $P_{T}^{X}>40$~GeV 
 & { 0 / 0.94 \small $\pm$0.11 } 
 & { 0 / 0.95 \small $\pm$0.12 } 
 & { { 1} / 0.07 \small $\pm$0.02 } 
 \\ 
\hline
\end{tabular} 
\end{table*}

\begin{table*}[t]
\caption{HERA II event yields in the search for isolated leptons with missing 
transverse momentum. The numbers are given for the electron and muon
channel, and after combination 
for different cuts $p_T^X$.}
\label{tab:iso_hera2}
\begin{tabular}{|c|c|c|c|}     
\hline 
{HERA~II 2003-2004}   & \multicolumn{3}{c|}{observed/expected}
\\ \hline \hline
{{\bf H1} ${\cal L}(e^+ p)=45$~pb$^{-1}$} & Electron & Muon &  Combined  \\ \hline
 Full sample 
 & { 7 / 4.08 $\pm$ \small 0.58  } 
 & { 1 / 1.2 $\pm$ \small 0.16 }  
 & { 8  / 5.28 $\pm$ \small 0.68 }  \\
 $P_{T}^{X}>25$~GeV 
 & { {\bf 3} / 0.74 $\pm$ \small 0.16 } 
 & { { 0} / 0.76 $\pm$ \small 0.11  } 
 & { 3 / 1.5 $\pm$ \small 0.24  } \\ 
\hline
\end{tabular}
\end{table*}

Events with isolated leptons and missing transverse momentum were
selected by requiring an isolated high $p_T$ lepton (electron, muon or tau) 
and missing transverse momentum.
For the remaining hadronic final state (jet) the transverse
momentum ($p_T^X$) is measured.
In the radial plane the reconstructed 
missing transverse momentum vector is required not to point into the
direction of the lepton or jet, which reduces genuine background from
deep inelastic scattering, and ensures that the missing transverse
momentum is due to an invisible particle ($\nu$).
The genuine SM ``background'' process is single W-production with the
leptonic decay. Searches have been performed by both experiments H1
and ZEUS in the decay channels into electrons and 
muons~\cite{ref:h1isol,ref:zeusisol}, and taus~\cite{ref:h1tau,ref:zeustau}.

In the HERA~I analyses 
good agreement between data
and the SM expectation
were found at both experiments, see
Table~\ref{tab:iso_hera1}. 
However, after applying a cut
$p_T^X>25$~GeV 11 electron and muon events were observed by the H1
experiment compared to an expectation of $3.44\pm0.59$, and
2 events  were observed in the tau channel by the ZEUS collaboration 
compared to an
expectation of $0.20\pm0.05$. 
Interestingly, neither of these excesses were confirmed by the
partner experiment. 

To enhance the limited statistics
 isolated lepton events were investigated in
the recent HERA~II dataset by the H1 collaboration~\cite{ref:hera2search}. 
The analyses were repeated in the electron and muon channel.
In total 8 new events were found compared to an expectation of $5.28\pm0.68$ 
events, see Table~\ref{tab:iso_hera2}.
For $p_T^X>25$~GeV 3 events, all found in the electron channel, survived
compared to an expectation of $0.74\pm0.16$ in the electron and
$0.76\pm0.11$ in the muon channel.

\begin{figure}[thb]
\setlength{\unitlength}{1cm}
\begin{picture}(10,5.8)
     \put(-0.1,0){\epsfig{file=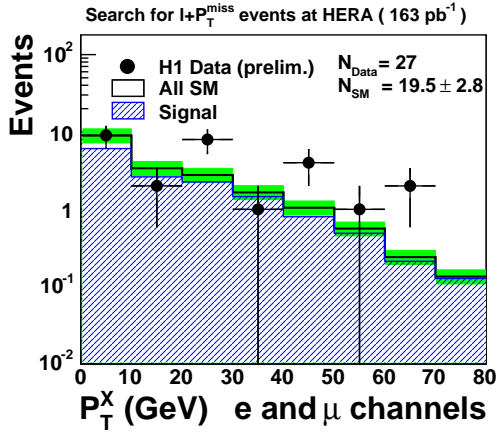,width=70mm}}
\end{picture}
\caption{The hadronic transverse momentum distribution in the electron
and muon channels combined (${\cal{L}}=163~pb^{-1}$)  compared to the
SM expectation (open histogram). The SM expectation is dominated by
real W production (hatched histogram). The total error on the SM
expectation is given by the shaded band.}
\label{fig:isol_ptx2}
\end{figure}

The distribution of the transverse momentum of the hadronic final
state of all H1 data combined corresponding to an integrated
luminosity of 163~pb$^{-1}$ is shown in Figure~\ref{fig:isol_ptx2}.
For $p_T^X>25$~GeV a clear excess of 14 events compared to $5.1\pm 1.0$
expected is visible.
However, more data and more detailed studies are required to resolve unambiguously  the
differences of event yields observed so far by the two collaborations H1
and ZEUS.

\section{Single top production} \label{sec:top}
The isolated lepton event class was discussed as possible signal for
anomalous production of singe top quarks at HERA
\cite{ref:theory_top} with the subsequent semileptonic decay
$t\rightarrow b \; \ell^+ \nu$.
As the SM expectation for the production of top quarks at HERA is
negligible this process provides an ideal testing ground for the
search for flavour changing neutral currents (FCNC).

\begin{figure}[th]
\setlength{\unitlength}{1cm}
\begin{picture}(10,6.7)
     \put(-0.3,-0.3){\epsfig{file=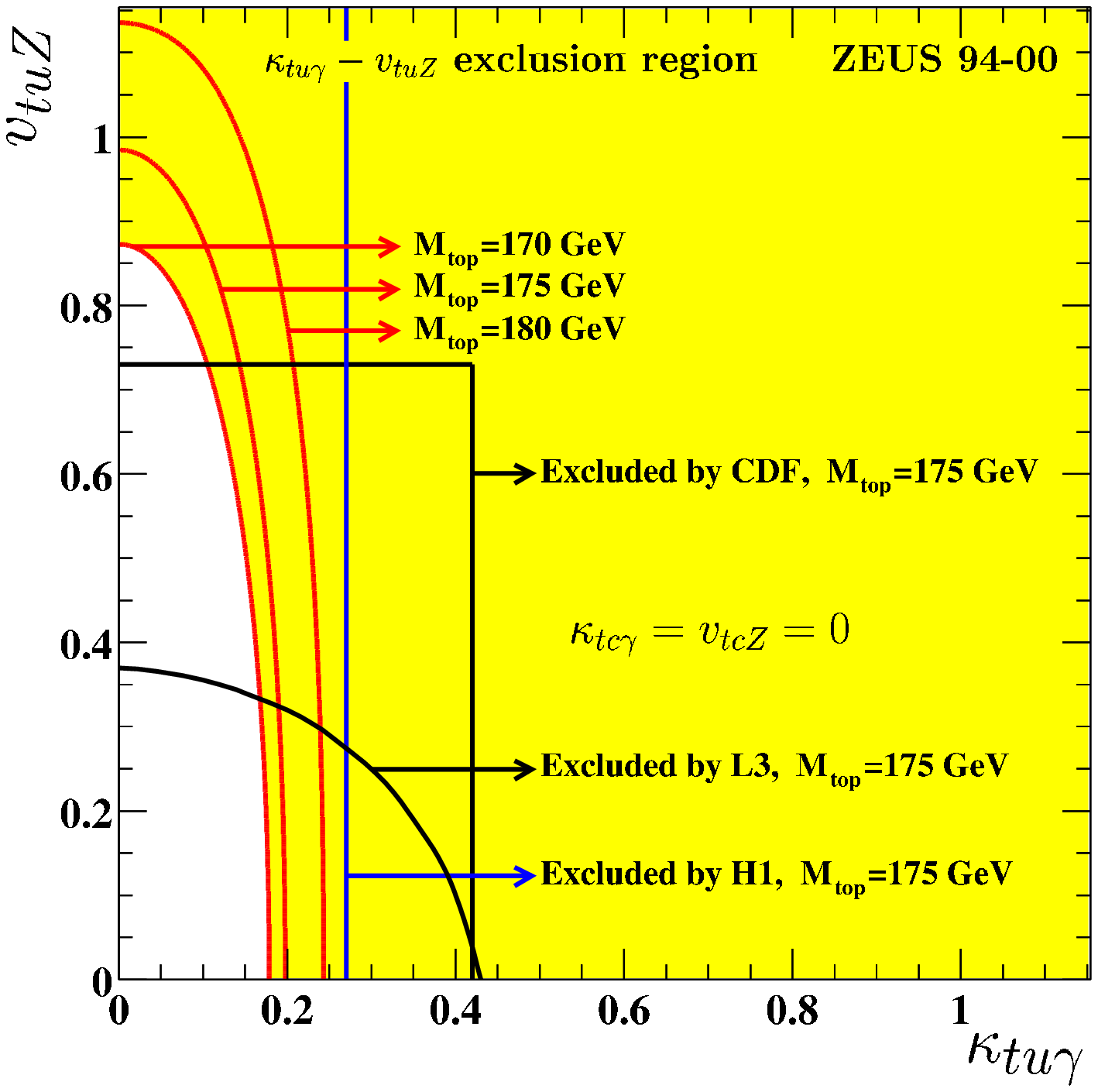,width=72mm}}
\end{picture}
\caption{Summary of exclusion limits at 95\% CL on the anomalous couplings
  $\kappa_{tu\gamma}-v_{tuZ}$ of the top quark. For details see text.}
\label{fig:limit_top}
\end{figure}

Isolated lepton events consistent with a top quark
decay signature were selected.
The main criteria is the
transverse momentum of the hadronic final state ($p_T^X$),
which can be associated to the b-jet and is
expected to be large. 
In order to gain sensitivity both experiments
\cite{ref:h1top,ref:zeusisol} have also studied the hadronic top decay, 
which is visible as three-jet events in both experiments.
The event yields measured by both experiments were found to be
 consistent with
the SM expectation and limits at the 95\% confidence level were set
on the top production cross section.

The cross section limits after combining 
the semileptonic and hadronic channels were found to be
$\sigma(ep\rightarrow etX)<0.43$~pb (H1) and
$\sigma(ep\rightarrow etX)<0.225$~pb (ZEUS).
Both limits can be directly converted into a
limit on the FCNC anomalous magnetic
coupling of the top quark to a $u$-quark and a photon:
$\kappa_{tu\gamma}<0.27$ (H1) and 
 $\kappa_{tu\gamma}<0.175$ (ZEUS).
$\kappa_{tu\gamma}$ can even be more constrained when allowing for
a non vanishing top coupling  to the Z-boson ($v_{tuZ}$),
see ZEUS limit in Figure~\ref{fig:limit_top}.
Also shown are limits obtained
at LEP from a search for single top production
and obtained at Tevatron from the study of radiative top decays.
The figure shows that HERA has an unique discovery potential for 
anomalous magnetic couplings of the top quark in a parameter space not
excluded by other experiments.

\section{Conclusion}
The $ep$ accelerator HERA has seen a major upgrade program to provide
higher luminosities and longitudinally polarised $e^\pm$ beams and first
data have been taken in $e^+p$ collisions. 
Recent results on a general model independent search, on a search for
isolated lepton events with missing transverse momentum, and on the
search for the anomalous production of single top quarks were
presented.
The trend of H1 at HERA~I detecting more isolated lepton events 
than predicted seems also to
continue at HERA~II in the electron channel.
At present, given the available statistics of isolated lepton 
events collected at the H1 and ZEUS
experiments, the results are inconclusive and clearly more data, which
are going to be collected in the upcoming years, will help to solve the puzzle.

\end{document}